\documentstyle[preprint,aps,epsf]{revtex}

\def\by#1#2{{\displaystyle {#1}\over \displaystyle {#2}}}
\def\d{{\rm d}}
\preprint {IMSc/2000/04/15}
\begin{document}
\title{Neutrinos from Stellar Collapse:\\
Comparison of the effects of three and four flavour mixings}

\author{Gautam Dutta, D. Indumathi, M. V. N. Murthy and G. Rajasekaran}

\address
{The Institute of Mathematical Sciences, Chennai 600 113, India.\\
}
\date{\today}
\maketitle
\begin{abstract}
We study the effect of non-vanishing masses and mixings among neutrino
flavours on the detection of neutrinos from stellar collapse by a water
Cerenkov detector. We consider a frame-work in which there are four
neutrino flavours, including a sterile neutrino, whose mass squared
differences and mixings are constrained by the present understanding of
solar, atmospheric and laboratory neutrino detection. We also include
the effects of high density matter within the supernova core. Unlike in
the three flavour scenario, we find that the number of events due to
the dominant process involving electron-antineutrinos changes
dramatically for some allowed mixing parameters.  Furthermore,
contributions from charged-current scattering off oxygen nuclei in the
detector can be considerably enhanced due to flavour mixing. We also
present a comparison between the two possible scenarios, namely, when
only three active neutrino flavours are present and when they are
accompanied by a fourth sterile neutrino.

\end{abstract}

\pacs{PACS numbers: 14.60.Pq, 13.15+g, 97.60.Bw}

\narrowtext

\section{Introduction}
In a recent paper \cite{dutta} (hereafter referred to as I), we
discussed in detail the signatures of 3 flavours of neutrinos from
stellar collapse.  The analysis was confined to Type II supernovae
(which occur when the initial mass of the star is between 8--20 solar
masses).  Based on the work of Kuo and Pantaleone \cite{KuoP}, where
they include mixing among all three neutrino flavours, we found that
the mixing between neutrino flavours leaves non-trivial signatures in
the detector. The main conclusion of the paper was that the effect of
mixing is to produce a dramatic increase in the events involving oxygen
targets\cite{haxton}. These will show up as a marked increase in the
number of events in the backward direction with respect to the forward
peaked events involving electrons as targets (more than 90\% of which
lie in a $10^\circ$ forward cone with respect to the supernova
direction for neutrinos with energies $E_\nu > 8$ MeV). In the absence
of any mixing, there will also be a few events in the backward
direction due to CC scattering on oxygen targets. Furthermore, the
observed $\overline{\nu}_e\,p$ events are the largest in number as well
as least sensitive to the mixing parameters within the three flavour
scenario. Hence they provide a direct test of the supernova models.
Since the angular distribution of these events is approximately
isotropic\cite{beacom}, they may be used to set the overall
normalisation.

The above analysis was done assuming the standard mass heirarchy
necessitated by the solar and atmospheric neutrino observations
\cite{snatm}.  In the meantime, several authors have looked at the
possible signatures of neutrinos and antineutrinos from supernova
collapse for Super Kamiokande and Sudbury Neutrino Observatory(SNO).
Dighe and Smirnov\cite{dighe} have looked at the problem of
reconstruction of neutrino mass spectrum in a three flavour scenario.
These authors as also Chiu and Kuo\cite{chiu} have compared the
signatures in the standard mass heirarchy and inverted-mass heirarchy.
While these papers and I incorporate the constraints from solar and
atmospheric neutrino observations, the important question of the mass
limits that may be obtained from the observation of time delay has also
been analysed by Choubey and Kar\cite{choubey} (see also the review by
Vogel\cite{vogel}). For a recent review which also discusses aspects of
locating a supernova by its neutrinos in advance of optical
observation, see ref.\cite{beacom}.

In this paper, we continue the main theme of paper I, and extend the
analysis to include a fourth sterile neutrino as well as discuss some
aspects of three flavour mixing not discussed in detail in I. The main
motivation to extend the analysis comes from the fact that such a
sterile neutrino is probably required in order to explain the recent
results from the LSND collaboration\cite{lsnd}. It is well known that
an explanation of the solar and atmospheric neutrino puzzles involve
two very different scales of mass-squared differences. As such, both of
them cannot be explained unless one invokes at least three neutrino
flavours.  However, the scale required in order to understand the
results from the LSND collaboration is different from both atmospheric
and solar neutrino puzzles. It has therefore become necessary to
introduce an extra-neutrino species, $\nu_s$, which however must remain
sterile due to the LEP constraint on the Z-width.

The main feature of the analysis as emphasised in I---that the charged
current (CC) events on oxygen nuclei (which show a preference to be in
the backward direction)---is preserved both in three and four flavour
scenarios. We also find that the signatures for three and four flavours
are very distinct for some parameter ranges in the dominant isotropic
events caused by electron antineutrino interactions with protons in
water.

The analysis presented in this paper follows closely the earlier
analysis presented in I and therefore we will only reproduce the main
outline and refer to I for some details.  In Sect. II we give an
outline of the framework and the mixing matrix as also matter effects
on mixing. We then use this to obtain expressions for the fluxes
reaching the detector in Sect. III. Sect. IV contains an analytic
discussion of the signatures in the case of extreme adiabatic and
extreme non-adiabatic mixing. In Sect. V we write down expressions for
the event rates due to interactions between different neutrino flavours
and scattering targets in the detector. Numerical results are presented
in Sect. VI both for the spectrum and the integrated number of events.
We present a summary and list our conclusions in Sect. VII. Appendix A
gives details of the choice of mixing matrix that we have used in the
paper. 

\section{Mixing in the presence of highly dense matter}

We briefly discuss mixing among four flavours of neutrinos (or
antineutrinos) and compute the neutrino survival and conversion
probabilities. 

Unlike in the three flavour scenario, the addition of a sterile
component forces us to prescribe a heirarchy of states in the mass
eigenstate basis. If neutrino oscillation is the mechanism for the
result from LSND, it is clear that the conversion of $\nu_{\mu}
\rightarrow \nu_e$ is governed by a mass scale in the range of 0.1
eV$^2$ to 1 eV$^2$. As shown in Fig.1, we choose two doublets separated
by this mass scale. In the lower doublet the mass-squared difference is
given by the appropriate scale for the solution of the solar neutrino
puzzle ($ \delta m^2_{S} < 10^{-5}$~eV${}^2$). Analogously, in the
upper doublet it is the atmospheric neutrino mass scale($ \delta
m^2_{ATM} \approx 10^{-3}$~eV${}^2$) that plays the crucial role. In
principle the sterile neutrino may be in either of these doublets.
However, if one believes that the atmospheric neutrino solution is
driven by the conversion of $\nu_{\mu} \rightarrow \nu_{\tau}$, then
the sterile neutrino should be in the lower doublet accounting for the
solar neutrino deficit.  While we take the sterile neutrino in the
lower doublet, the final results are not crucially dependent on where
we position the sterile neutrino as long as the two doublet scheme is
adhered to.

The mixing matrix, which relates the flavour and mass eigenstates in
the four flavour scenario has six angles and the CP-violating phases
which we do not consider here. As is the convention, we denote the
mixing angle in the lower doublet by $\omega$ and the mixing angle in
the upper doublet by $\psi$. The resulting complicated mixing matrix
involving six mixing angles is greatly simplified by application of the
{\sc chooz} \cite{chooz} constraint. The {\sc chooz} result implies
that $\sin \theta \le \epsilon$, where $\theta$ is either the (13) or
(14) mixing angle and $\epsilon \le 0.16$. We will therefore replace
these angles by their maximum possible values allowed by {\sc chooz}.
Furthermore, we choose $\sin \theta_{13} \sim \sin \theta_{14} \sim
\sin \theta_{23} \sim \sin \theta_{24} \sim \epsilon$. (The oscillation
probabilities we are mainly interested in depend on the $\nu_e$ survival
and conversion probabilities. These involve only $\theta_{13}$
and $\theta_{14}$ mixing angles. As such $\theta_{23}$ and
$\theta_{24}$ may be kept arbitrary and have been fixed to $\epsilon$
for convenience. Note that the {\sc chooz} constraint corresponds to
the replacement $\sin \phi \to \epsilon$ in the three flavour case
analysed in I, where $\phi$ is the (13) mixing angle to which the {\sc
chooz} constraint applies. This choice of hierarchy and mixings is
consistent with already known data from various laboratory and
atmospheric neutrino experiments.  For more details on the constraints
on mass squared differences and mixing angles from various experiments,
see Appendix A. Then the flavour eigenstates are related to the four
mass eigenstates in vacuum (for both neutrinos and antineutrinos)
through a unitary transformation,
\begin{equation}
\left[ \begin{array}{c} \nu_e \\ \nu_s \\ \nu_{\mu} \\ \nu_{\tau}
\end{array} \right] = U^v 
\left[ \begin{array}{c} \nu_1 \\ \nu_2 \\ \nu_3 \\ \nu_4 
\end{array} \right],
\end{equation}
where the superscript $v$ on the r.h.s. stands for vacuum.  Within the 
two doublet scheme the $4 \times 4$ unitary matrix, $U^v$, may be written as
\begin{equation}
U^v = \left( \begin{array}{cccc}
      (1-\epsilon^2)c_{\omega} & ~~~(1-\epsilon^2)s_{\omega} & \epsilon 
&\epsilon \\
     -(1-\epsilon^2)s_{\omega}-2\epsilon^2 c_{\omega} & 
(1-\epsilon^2)c_{\omega} -2\epsilon^2 s_{\omega}& \epsilon &\epsilon \\
\epsilon(s_{\omega}-c_{\omega})(c_{\psi}+s_{\psi})& 
-\epsilon(s_{\omega}+c_{\omega})(c_{\psi}+s_{\psi})& 
     (1-\epsilon^2)c_{\psi}-2\epsilon^2 s_{\psi} & 
     (1-\epsilon^2)s_{\psi}\\ 
\epsilon(s_{\omega}-c_{\omega})(c_{\psi}-s_{\psi})& 
-\epsilon(s_{\omega}+c_{\omega})(c_{\psi}-s_{\psi})& 
     -(1-\epsilon^2)s_{\psi}-2\epsilon^2 c_{\psi} & 
     (1-\epsilon^2)c_{\psi}\\ 
      \end{array} \right),
\label{eq:vac}
\end{equation}
where $s_{\omega} = \sin \omega$ and $c_{\omega} = \cos \omega$, etc. The
angles $\omega$ and $\psi$ can take values between $0$ and $\pi/2$.
The mixing matrix given above is unitary up to order $\epsilon^2$. Since 
$\epsilon$ is a small parameter, we do not need to go beyond this order. 
Later on we will show that the survival and oscillation probabilities that 
we need are such that they do not depend on the mixing in the upper 
doublet, namely the angle $\psi$. For all practical purposes it can be 
set to any value, in particular, zero. 

The masses of the eigenstates in vacuum are taken to be $\mu_1$,
$\mu_2$, $\mu_3$ and $\mu_4$.
In the mass eigenbasis, the $({\rm mass})^2$ matrix is diagonal:
\begin{eqnarray}
M_0^2   & = & \mu_1^2 I\!\!I + \left( \begin{array}{cccc}
			 0 & 0 & 0 &0 \\
			 0 & \delta_{21} & 0 &0\\
			 0 & 0 & \delta_{31} &0\\
			 0 & 0 & 0 &\delta_{41} \\
			 \end{array} \right),  \nonumber \\
 & = & \mu_1^2 I\!\!I + \Delta M_0^2~,  
\end{eqnarray}
where the mass squared differences are given by $\delta_{21} = \mu_2^2
- \mu_1^2$, $\delta_{31} = \mu_3^2 - \mu_1^2$ 
and $\delta_{41} = \mu_4^2 - \mu_1^2$. 

Without loss of generality, we can take $\delta_{21}$, $\delta_{31}$
and $\delta_{41}$ to be greater than zero; this defines the standard
hierarchy of masses consistent with the range of the mixing angles, as
specified above.  Neutrino oscillation amplitudes are independent of
the first term so we drop it from further calculation. In the flavour
basis, therefore, the relevant part of the mass squared matrix has the
form,
\begin{equation}
\Delta M_v^2   =  U^v \, \Delta M_0^2 \, {U^v}^{\dagger}. 
\label{eq:mass}
\end{equation}
where $U^v$ is the mixing matrix in vacuum.

\subsection{Matter effects for neutrinos}
The relevant matter effects may  be included by a modified mass squared
matrix,
\begin{equation}
\Delta M_m^2 = \Delta M_v^2 + \Delta M_A,
\label{eq:mmsq}
\end{equation}
where the matter effects are included in,
\begin{equation}
\Delta M_A = \left( \begin{array}{cccc}
A_1(r)& 0 & 0 & 0 \\ 0 & A_2(r) & 0 & 0 \\ 
0 & 0 & 0 & 0 \\ 
0 & 0 & 0 & 0 \\ 
       \end{array} \right)~,
\label{eq:A}
\end{equation}
with the $A_i(r)$ given by
\begin{equation}
A_1(r) = \sqrt{2}~ G_F ~N_e (r) \times 2 E~, \label{eq:defA1}
\end{equation}
and
\begin{equation}
A_2(r) \approx \sqrt{2}~ G_F ~N_n (r) \times  E~, \label{eq:defA2}
\end{equation}
which are  proportional to the electron number density, $N_e(r)$, and
neutron number density $N_n(r)$ respectively. We set $N_e(r) \simeq f_e
N_n(r) $, where $f_e$ is the electron fraction which is less than one,
in the supernova core. Here $r$ is the radial distance from the centre
of the star. Note that we have subtracted the neutral current
contribution to the active flavours in the interaction term. Since the
sterile neutrino does not interact at all, we have added and subtracted
the neutral current contribution; hence the appearance of $A_2$ in the
mass-squared matrix in matter. The detailed modifications due to matter
effects are discussed in Appendix A of I.

As noted in I, the maximum value of $A_1$ occurs at the core and is
approximately $2\times10^{7} E \ {\rm eV}^2$, where $E$ is the neutrino
energy in MeV. $A_2$ is somewhat smaller. The modification due to the
matter dependence is similar to the case of solar neutrinos, although,
unlike in the case of solar neutrinos, all active flavours are produced
in the supernova core.

It is clear that the mass squared matrix is no longer diagonalised by
$U^v$ in the presence of matter; we therefore diagonalise $\Delta
M_m^2$ in order to determine the matter-corrected eigenstates. The
value of $A_i$ for energetic neutrinos (of a few MeV to tens of MeV) in
the core is several orders of magnitude greater than these mass-squared
differences.  The eigenvalue problem may thus be solved perturbatively,
with the following hierarchy:  $A_i (\hbox{core}) \gg \delta_{41} >
\delta_{31} \gg \delta_{21}$. As a result, the electron neutrino
undergoes three Mikheyev, Smirnov and Wolfenstein (MSW) resonances
\cite{MSW}. In Fig. 2 we show schematically the level crossing pattern
for neutrinos in the presence of matter. Note that in this case there
are several level crossings because of the presence of the sterile
neutrino.

\subsubsection{The adiabatic case}

Following the method outlined in I, we compute the eigenstates at
production (almost all the active neutrino flux is produced in the
core).  Unless $\epsilon$ is extremely small, the propagation is
adiabatic as may be shown by computing the Landau-Zener(LZ)
probabilities at each resonance. While we include LZ probabilities in
the actual numerical computation, we will first state the relevant
probabilities in the case of adiabatic propagation of neutrinos
produced in the supernova core.  The average transition probability
between two flavours $\alpha$ and $\beta$ is denoted by $P_{\alpha
\beta}$ where $\alpha,\beta = e, \mu, \tau, s$. These are obtained by
observing that almost all the $\nu$ flux is produced in the highly
dense core. Here $A_i \approx 10^{7}$ eV${}^2$, so one may take the
extreme limit in which the density is infinite. Then the electron
neutrino is produced as the highest mass eigenstate, that is,
$$
|\nu_e\rangle^m = |\nu_4\rangle^m.
$$
A similar conclusion holds for the sterile partner in the lower
doublet. In the case of adiabatic propagation the mass eigenstates
produced in matter remain the same in vacuum. Therefore to a good
approximation, the probabilities (of survival or transition into each
other) for these two flavours, $\nu_e$ and $\nu_\mu$, are just the
overlap of the highest two mass eigen states with the corresponding
vacuum eigen states, and are small, as can be seen from the
relevant entries in the matrix, $U^v$, in eq. (\ref{eq:vac}):
\begin{equation}
P_{ee}=P_{es}=P_{se}=P_{ss}=\epsilon^2~,
\end{equation}
independent of other mixing angles and the mass squared differences. Note
that $\epsilon^2$ is constrained by the limit set by the {\sc chooz}
collaboration and is small. Indeed in the purely adiabatic transition
therefore the survival probability of the electron type neutrino is rather
small. We also note that it is sufficient to know only these 
probabilities as long as the $\nu_{\mu}$ and $\nu_{\tau}$ are not 
separately detected. Furthermore, note that there is no initial flux for 
the sterile neutrino, but it can arise after oscillation due to mixing
from other flavours.  

\subsubsection{The non-adiabatic case}

A general non-adiabatic case is harder to discuss analytically. Since
there are many crossings, one has now to consider non-adiabatic
transitions at all these crossings. Because of the parametrization it
is easy to see that non-adiabatic effects are introduced as a result of
the values chosen for $\epsilon$ and $\omega$.  The value of $\epsilon$
determines whether non-adiabatic jumps are induced at the upper
resonances while the value of $\omega$ determines whether the
non-adiabatic jump occurs at the lower resonance.  This statement holds
both for three and four flavours since in both cases the
non-adiabaticity in the upper resonance(s) is controlled by $\epsilon$,
apart from mass squared differences.

Note that for a large range of $\epsilon$, allowed by the {\sc chooz}
constraint, the evolution of the electron neutrino is adiabatic. As a
result the lower resonance does not come into the picture at all except
when $\epsilon$ is very close to zero, where the jump probability
abruptly changes to one. \cite{KuoP}The subsequent discussion is therefore
relevant only when $\epsilon$ is almost vanishing which is not ruled
out by the known constraints except in LSND.  
The extreme non-adiabatic case occurs
when the nature of transitions at the lowest resonance which involves
the crossing of the first and the second mass eigenstates is
nonadiabatic. Partial non-adiabaticity results when the transition in
the lowest resonance is adiabatic while in the upper resonances it is
non-adiabatic.  The probability of jump at the lower resonance is given
by $P_L$ which is in general a function of $\omega$ and the mass
squared difference $\delta m_{12}^2$.  In our calculations we have used
the form discussed in the appendix of I (see also \cite{LZ}).

The relevant survival and oscillation probabilities for electron-neutrinos 
in the non-adiabatic case are given by, 
\begin{eqnarray}
P_{ee}&=& (1-2 \epsilon^2)[(1-P_L)\sin^2 \omega~ + P_L \cos^2\omega],\\ 
P_{es}&=& (1-2 \epsilon^2)[(1-P_L)\cos^2 \omega~ + P_L \sin^2\omega],\\ 
P_{ss}&=& P_{ee} + 2 \epsilon^2 (1-2P_L) \sin 2\omega ,\\ 
P_{se}&=& P_{es} - 2 \epsilon^2 (1-2P_L) \sin 2\omega ,
 \end{eqnarray}
where $\omega$ is as usual the vacuum mixing angle defined earlier. In
the three flavour case the only relevant probability is,
\begin{equation}
P_{ee}= (1 - \epsilon^2)[(1-P_L)\sin^2 \omega~ + P_L \cos^2\omega].
\end{equation}
Note that, since $\epsilon$ is small, the flux at the detector is 
entirely controlled by $\omega$ and $P_L$ which is also a function of 
$\omega$. 

\subsection{Matter effects for antineutrinos}
We now consider the case of $\overline{\nu}_e$ propagation in highly
dense matter. The only change in this case is that the matter dependent
term in the relevant part of the mass squared matrix has the opposite
sign (to that in Eq.~(\ref{eq:defA1})), that is,
\begin{equation}
A_i(r) = -\sqrt{2}~ G_F ~N_i (r) \times 2 E~; N_1= N_e,~N_2=N_n/2.
\end{equation}  
The analysis goes through as in the case of $\nu_e$ propagation through
matter.
There are no Landau-Zener jumps to consider in this case
since the resonance conditions are never satisfied unless the mass
hierarchy is altered.  The propagation is therefore adiabatic and the
survival probability is obtained by simply projecting the $\vert
\overline\nu_1\rangle $ eigenstate on to the flavour eigenstate in
vacuum (at the detector). The antineutrino survival and transition
probabilities for $e$ and $s$ flavour neutrinos are,
\begin{eqnarray}
P_{\overline{e}\overline{e}} &=& (1-2 \epsilon^2)\cos^2\omega~,\\ 
P_{\overline{e}\overline{s}} &=& (1-2 \epsilon^2)\sin^2\omega~,\\ 
P_{\overline{s}\overline{e}} &=& (1-2 \epsilon^2)\sin^2\omega +
		      2\epsilon^2 \sin 2\omega~, \\
P_{\overline{s}\overline{s}} &=& (1-2 \epsilon^2)\cos^2\omega -
		      2\epsilon^2 \sin 2\omega~,
\end{eqnarray}
where $\omega$ is as usual the vacuum mixing angle defined 
earlier. As before, we do not need to know the other probabilities for
our analysis.

\section{Fluxes at the detector}

We briefly compute the neutrino flux at the detector in the presence of 
mixing and compare the three and four flavour scenarios. 

Following Kuo and Pantaleone \cite{KuoP}, we denote the flux
distribution, ${\rm d} \phi_\alpha^0 /{\rm d} E$, of a neutrino (or
antineutrino) of flavour $\alpha$ with energy $E$ produced in the core
of the supernova by $F_\alpha^0$. In particular we use the generic
label $F_x^0$ for flavours other than $\nu_e$ and $\overline\nu_e$
since
\begin{equation}
F_x^0 = F_{{\mu}}^0
= F_{\overline{\mu}}^0
= F_{{\tau}}^0
= F_{\overline{\tau}}^0~.
\end{equation}
All these flavours are produced via the neutral-current (NC) pair
production processes and therefore have the same flux for all practical
purposes. However, the $\nu_e$ and $\overline\nu_e$ fluxes are
different from each other and the rest since they are produced not only
by pair production but also derive contribution from charged-current
(CC) processes. Note that there is no production of sterile neutrinos
at source.

The flux reaching the detector from a supernova at a distance $d$ from
earth is reduced by an overall geometric factor of $1/(4\pi d^2)$.
Apart from this, there is a further modification of the observed flux
due to oscillations in the presence of matter. 
The flux on earth, in the various flavours, is
given in terms of the flux of neutrinos produced in the core of the
supernova by,
\begin{eqnarray}
F_{e} & = & P_{ee} F_{e}^0 +P_{e\mu}F_{{\mu}}^0 + 
                P_{e\tau}F_{{\tau}}^0, \nonumber \\ 
      & = & P_{ee}F_{e}^0 +(1-P_{ee}-P_{es})F_x^0~, 
\label{fe}
\end{eqnarray}
where we have made use of the constraint $\sum_{\beta} P_{\alpha\beta} =1$.
Since $\nu_{\mu}$ and
$\nu_{\tau}$  induced events cannot be separated in water Cerenkov
detectors, their combined flux on earth may be written as
\begin{eqnarray}
2 F_x & = & F_{{\mu}} + F_{{\tau}}~, \nonumber \\ 
      & = & (P_{ee}+P_{es}+P_{se}+P_{ss})F_x^0 
                +(1-P_{ee}-P_{se})F_{e}^0~,
\label{fx} 
\end{eqnarray}
and
\begin{equation}
F_s = (1-P_{ss}-P_{se})F_x^0 + P_{se} F_{e}^0~.
\label{fs} 
\end{equation}
Note that a part of active neutrino flavours are converted into sterile 
due to mixing. The total flux is however conserved since,
$$
F_{e} + 2F_x + F_s = F_{e}^0 + 2F_x^0~.
$$

For comparison, the three-flavour case may be obtained by setting all
the conversion probabilities involving the sterile neutrino to zero,
and setting $P_{ss} = 1$.

\section{Analytical results for fluxes}
Without any numerical estimation, or choice of supernova model, it is
possible to analyse the adiabatic case as well as the extreme
nonadiabatic case and obtain gross features of the mixing for both 3 and
4 flavour mixings. We shall analyse the mixings in the neutrinos and
antineutrinos separately. We begin with the adiabatic case.

\subsection{Neutrino fluxes: Adiabatic case}
For adiabatic propagation, Landau Zener jumps must be
negligible. This in turn means that the larger the mass difference, the
smaller the mixing angle one can accomodate. We shall assume that this is
satisfied for the case under consideration here.

The results for the neutrino fluxes are given in
Table~\ref{tab:adnu}. It is seen that the fluxes are independent of the
(12) mixing angle, $\omega$, in all cases.

\subsubsection{4-flavours} For the 4-flavour case, it is striking that
the electron neutrino flux and the sterile neutrino flux (not
observable in the detector) are equal and completely deplete the
original spectrum in $\nu_{\mu,\tau}$.  The $\nu_{\mu,\tau}$ flux at
the detector is entirely made up by almost complete conversion of
electron neutrinos.  From the supernova models, it is reasonably well
known that the spectrum of $\nu_e$ is colder than $\nu_{\mu, \tau}$. In
effect therefore the mixing interchanges the hot and the cold spectra
in the supernova, with about half of the hotter spectrum lost in the
sterile neutrinos.

\subsubsection{3-flavours} The 3-flavour $\nu_e$ flux is not very different
from the 4-flavour one: about half of the hotter spectrum is lost and
reappears as the electron neutrino spectrum in the detector (see
Table~\ref{tab:adnu}). There is no sterile neutrino; hence the
$\nu_{\mu,\tau}$ spectra are not further depleted, in contrast to the
4-flavour case. However, this difference in $\nu_{\mu,\tau}$ spectra
between 3 and 4 flavour mixing is not easily observed since water Cerenkov
detectors see very few events from flavours other than electron type
neutrinos. The evidence for physical loss of flux into a sterile
channel may be observed from neutral current events in detectors such
as SNO.

In both 3 and 4 flavour cases, there is mixing into the $\nu_e$ spectrum
of the hotter $\nu_x$ spectrum. As we have seen in I, this admixture of
hotter spectrum is signalled by a strong increase in the number of
backwardly peaked events due to the opening up of the $\nu \, O$ charged
current channel in the detector, and should be a clean indicator of
$\nu_e$-$\nu_x$ mixing.

\subsection{Antineutrino fluxes}

The basic analysis is similar to the case of neutrino fluxes given in
equations (\ref{fe}),(\ref{fx}) and (\ref{fs}). The changes occur only
at the level of substituting for the appropriate probabilities, given
in the previous section. These are different for neutrinos and
antineutrinos.

Unlike in the neutrino sector, predictions in the anti-neutrino sector
require inputs on $\omega$. Within the schemes considered in this
paper, this is obtained only from an analysis of the solar neutrino
deficit. Then there are two possible solutions for $\sin \omega$---the
so-called small mixing angle (SMA) MSW solution, and also the large
mixing angle (LMA) solutions. The LMA solutions arise both with matter
effects included and in the so-called just-so solutions where there are
no matter effects. At present, all three scenarios are consistent with
the solar neutrino data although the mass-squared difference required
in these cases is vastly different---$\delta m_{12}^2 \approx 10^{-5}$
eV${}^2$ in the case of MSW solution with LMA, $\delta m_{12}^2 \approx
10^{-7}$ eV${}^2$ for LMA with vacuum (LMA-V) and $\delta m_{12}^2
\approx 10^{-6}$ eV${}^2$ for SMA solution.

In any case, whatever the value of $\omega$, the propagation of
antineutrinos always remains adiabatic (unless we choose an inverted
mass hierarchy, which we do not consider here).  The results for
arbitrary $\omega$ are shown in Table~\ref{tab:adnubar}. While there is
hardly any admixture of the hotter $\overline\nu_{\mu,\tau}$ spectrum into
$\overline{\nu}_e$ (independent of $\omega$) in the 4-flavour case, the
extent of this mixing depends on $\omega$ in the 3-flavour case.

\subsubsection{Small $\omega$} It is easy to see from Table~\ref{tab:adnubar}
that for small $\omega$, the $\overline{\nu}_e$ flux is almost
unaltered, in both 3 and 4-flavour mixing cases. These are then
indistinguishable (and also indistinguishable from the no-mixing case).

\subsubsection{Large $\omega$} In 4-flavour mixing, the original
$\overline{\nu}_e$ flux is depleted by a factor of $\cos^2\omega$,
while still receiving no contribution from $\nu_{\mu,\tau}$ (See
Table~\ref{tab:adnubar}).
This depletion is seen at all energies and should be easily
observable since the CC process $\overline{\nu}_e \, p \to  e^+ n$ has
by far the largest cross-section in the detector; furthermore, it is
approximately isotropic in distribution \cite{beacom}.  In contrast, a
large value of $\omega$ increases the contribution of $\nu_{\mu,\tau}$
in $\overline{\nu}_e$ while proportionately decreasing the survival of
the original $\overline{\nu}_e$ in 3 flavours. Such a mixing will be
observable at the higher energy end of the spectrum, where, however,
the event rates are low. Hence, the behaviour of the isotropic events
will be a clean signal of the number of flavours, provided, of course,
that the original fluxes are accurately known. This depends on the
models of stellar collapse. However, it may be possible to normalise
the overall spectrum from the initial burst neutrinos (which are purely
$\nu_e$ type and give a forward electron distribution in the Cerenkov
detector).

The $\overline{\nu}_x$ flux is essentially unaltered, independent of
$\omega$ in the case of 4 flavours. With 3 flavours, the original
$\overline{\nu}_x$ flux is depleted, and is compensated for by a
proportional contribution from the $\overline{\nu}_e$ flux. Hence the
results are reversed compared to the $\overline{\nu}_e$ case. This
signal, however, may be difficult to observe as there are very few
events in this channel.

\subsection{Sterile neutrino fluxes}

We notice from Table~\ref{tab:adnu} and Table~\ref{tab:adnubar} that in
the four flavour case, there is actual loss of flux into the sterile
channel. Independent of $\omega$, half of the hot $\nu_{\mu,\tau}$
spectrum is lost into the $\nu_s$ channel, while a portion proportional
to $\sin^2 \omega$ of the $\bar\nu_e$ flux is lost into the $\bar\nu_s$
channel. From a detection point of view, this change, even if
dramatically large, cannot be observed, unless through observation of
neutral current events.

\subsection{Neutrino fluxes: Non-adiabatic case}

As we have seen, the various survival and transition probabilities in
the non-adiabatic case depend on both $\epsilon$ and $\omega$. For a
large range of $\epsilon$ the propagation is adiabatic. Hence
non-adiabatic effects are important only for small values of $\epsilon$
when the jump probability at the upper resonances abruptly changes to
one. Then, the flux in this case is driven entirely by
$\omega$, which also determines the size of the Landau-Zener jumps at
the lower resonance through the value of the jump probability, $P_L$.

\subsubsection{Small $\omega$} The neutrino fluxes at the detector,
calculated using these probabilities, are summarised in
Table~\ref{tab:nonad}. For comparison we have also given the results
for three flavours. It is clear from the table that if the mixing
angle $\omega$ is small, the $\nu_e$ flux at the detector is
approximately given by,
\begin{eqnarray}
F_e & \approx & P_L F_e^0 + (1-P_L) F_x^0~, (3\hbox{-flavours}),\\
F_e & \approx & P_L F_e^0~ \qquad \qquad \qquad (4\hbox{-flavours})~.
\end{eqnarray}
Hence the signals in the 3 and 4 flavour cases are drastically
different: there is no mixing of the hot spectrum into the original
$\nu_e$ spectrum in the 4 flavour case. 

\subsubsection{Large $\omega$} In the case of maximal mixing,
$\omega=\pi/4$, we have
\begin{eqnarray}
F_e & \approx & \frac{1}{2}[ F_e^0 + F_x^0]~, (3\hbox{-flavours}), \\
F_e & \approx & \frac{1}{2}F_e^0 \qquad \qquad \qquad (4\hbox{-flavours})~.
\end{eqnarray}
Near maximal mixing, the LZ jump probability $P_L$ plays no role at
all.  However, the $\nu_e$ flux at the detector again differs
dramatically from three to four flavours (See Table~\ref{tab:nonad}).
The three flavour result is also unlike the fully adiabatic case since
we have an equal mixture of the cold and the hot spectrum due to
mixing. Furthermore, the four flavour scenario differs from the four
flavour adiabatic case quite dramatically.  Unlike the adiabatic case,
here there is no contribution from the hot spectrum, while the cold
spectrum is depleted by about half.

In short, the $\nu_e$ signal in the nonadiabatic case is very different
not only between 3 and 4 flavours, but also is very different from the
adiabatic case. Here we have discussed the two extreme cases where the
lower resonance undergoes or does not undergo Landau Zener jumps. We
clarify the general case numerically below.

In the case of anti-neutrinos, non-adiabatic transitions do not occur
since there are no level crossings with the mass hierarchy assumed in this
analysis. However, in the case of an inverted mass hierarchy that allows
for non-adiabatic proagation in the antineutrino sector, the survival
probabilities are identical to those given above.  We will not discuss
this situation further here. 

In the next section, we compute numerically the fluxes, for some 
preferred values of $\omega$ and $\delta m_{12}^2$ (obtained from an
analysis of solar neutrino data) with the jump 
probability determined by these parameters. We will also predict event
rates and the total number of events in the different cases we have
discussed above.

\section{Cross-sections and event rates}

The basic quantity we are interested in is the distribution of events
in the detector as a function of the energy of the detected particle.
For a detailed discussion, for example, see I. Since we are interested
in water Cerenkov detectors, the relevant interactions are,
\begin{eqnarray} \nonumber
\nu_\ell (\overline{\nu}_\ell) + e^- & \to & 
\nu_\ell (\overline{\nu}_\ell) + e^-~, \ell = e, \mu, \tau~; 
\usecounter{reaction} \label{r1} \\
\overline{\nu}_e + p & \to & e^+ + n~;
\usecounter{reaction} \label{r2} \\ 
\nu_e + {}^{16} O & \to & e^- + {}^{16} F~, \\
\usecounter{reaction} \label{r3}
\overline{\nu}_e + {}^{16} O & \to & e^+ + {}^{16} N~.
\usecounter{reaction} \label{r4}
\end{eqnarray}
The oxygen cross-sections have been taken from Fig.~1 of Haxton
\cite{haxton}. All other cross-sections are well known. As the
interactions on protons and oxygen nuclei are purely CC interactions,
they involve only $\nu_e$ and $\overline{\nu}_e$. The interaction with
electrons involves both CC and NC interactions for $\nu_e$ and
$\overline{\nu}_e$ and only NC interactions for all other flavours. The
$\overline{\nu}_e p$ cross-section is the largest, being proportional
to the square of the antineutrino energy. In terms of total number of
events, therefore, water Cerenkov detectors are mostly dominated by
$\overline{\nu}_e$ events.  However, the different interactions in the
detector have distinct angular distributions: The elastic
neutrino-electron cross-sections are forward peaked, especially for
neutrinos with energies \raisebox{-0.1cm}{$\stackrel{\displaystyle
>}{\sim}$} 10 MeV \cite{Bahcallb,GM}, while the proton cross-section is
approximately isotropic in the lab frame. There is a slight excess of
backward events for energies below 15 MeV, and slight excess in the
forward direction at higher energies\cite{beacom} but the excess in
either direction at the relevant energies are limited to few percent of
the total number of events. Finally, the CC $\nu_e$
($\overline{\nu}_e$) cross-section on oxygen, although having a rather
large threshold of 15.4 MeV (11.4 MeV) \cite{haxton}, increases rapidly
with incoming neutrino energy and the number of events in the backward
direction increases substantially with energy of the incoming neutrino
above the threshold.

The time integrated event rate, from neutrinos of flavour $\alpha$ and
energy $E$, as a function of the recoil electron (or positron) energy,
$E_e$, is as usual given by,
\begin{equation}
\by{\d N_\alpha^{t,p}(E_e)}{\d E_e} = \by{n_t}{4\pi d^2} \sum_b \Delta
                   t_b \int \d E F_\alpha (b) \by{\d
		   \sigma_p}{\d E_e}~.
\label{eq:rate}
\end{equation}
Here the index $b$ refers to the time interval\footnote{The constant
factor $\Delta t_b$ should also appear in the corresponding equation,
Eq.~(27), in I.}. Note that the flux distribution, $F_\alpha (b)$
includes the effects of mixing in the hot dense core and is a function
of the time-dependent temperature, $T_b$. The index $p$ refers to any
of the various processes through which the neutrino $\alpha$ can
interact with the target, $t$, in the detector. Here $n_t$ refers to
the number of scattering targets (of either $e$, $p$ or ${}^{16}O$)
that are available in the detector. The total number of events from a
given flavour of neutrino in a given bin, $k$, of electron energy
(which we choose to be of width 1 MeV) then is
\begin{equation}
N_\alpha^{t,p}(k) = \int_k^{k+1} \d E_e \by{\d N_\alpha^{t,p}}{\d E_e}~.
\end{equation}
In the next section we give detailed numerical estimates both for the
spectrum and the integrated events for some specific representative
values of $\epsilon$ and $\omega$, as constrained by other experiments.

\section{Results and Discussion}

As in I, we compute the time integrated event rate at a prototype 1 Kton 
water Cerenkov detector from neutrinos emitted by a supernova exploding 10
KPc away.  Results for any other supernova explosion may be obtained by
scaling the event rate by the appropriate distance to the supernova and
the size of the detector, as shown in Eq.~(\ref{eq:rate}).  We assume
the efficiency and resolution of such a detector to be perfect.

We use the luminosity and average energy distributions (as functions of
time) as given in Totani et al. \cite{Totani}, based on the numerical
modelling of Mayle, Wilson and Schramm \cite{BL}. In a short time
interval, $\Delta t_b$, the temperature can be set to a constant,
$T_b$. Then, the neutrino number flux is described, in this time
interval, by a thermal Fermi Dirac distribution,
\begin{equation}
F_\alpha^0(b) = N_0 \by{{\cal L} (\alpha)}{T_b^4}
                       \by{E^2}{(\exp(E/T_b) +1)} ~,
\end{equation}
for neutrinos of flavour $\alpha$ and energy $E$ at a time $t$ after
the core bounce. Here $b$ refers to the time-bin, $t = t_0 + b \Delta t
$.  Hereafter, we set the time of bounce, $t_0 = 0$. The overall
normalisation, $N_0$, is fixed by requiring that the total energy
emitted per unit time equals the luminosity, ${\cal L} (\alpha)$, in
that time interval.The total emitted energy in all flavours of
neutrinos is about $2.7 \times 10^{53}$ ergs, which is more or less
equally distributed in all flavours. The number of neutrinos emitted in
each flavour, however, is not the same since their average energies are
different.

It turns out that the results are not very sensitive to the time
dependence of the temperature profile. For instance, using
time-independent temperatures of 11, 16, and 25 MeV for the $\nu_e$,
$\overline{\nu}_e$ and the $\nu_x$ thermal spectra (instead of the
above time-dependent ones) changes the results by less than 5\% in the
antineutrino sector and by about 8\% in the neutrino sector. However the
difference between the average temperatures of the electron and other
types of neutrinos is crucial to the analysis.

In addition we impose the following known constraints on the mixing
matrix in vacuum both for three and four flavour scenarios. Consistent
with the {\sc chooz} constraint, namely $\epsilon \le 0.16$ (which we
have already imposed at the level of the parametrisation itself), we set
$\epsilon = 0.087$.  Furthermore, the following constraints derived
from solar and atmospheric neutrino observations are also taken into
consideration:

The constraint from the atmospheric neutrino analysis implies that the
relevant angle $\psi (\approx \pi/4)$, is near maximal and the relevant
mass squared difference is of the order of $10^{-3}$. Neither of these
constraints directly enter our calculations except to determine whether
the upper resonance is adiabatic or not depending on the value of
$\epsilon$ as constrained by the {\sc chooz} findings. We consider both
here.

The more relevant constraint follows from the solar neutrino physics. 
Here there are three possible best fits to the combined data on solar 
neutrinos\cite{Bahcallr}: 
\begin{enumerate}
\item $\sin^2(2\omega) = 6.0\times 10^{-3}, \delta m_{12}^2 = 5.4 \times 
10^{-6}$ eV${}^2 $ (SMA). The small angle MSW solution.
\item $\sin^2(2\omega) = 0.76, \delta m_{12}^2 = 1.8 \times 
10^{-5}$ eV${}^2 $ (LMA). The large angle MSW solution.
\item $\sin^2(2\omega) = 0.96, \delta m_{12}^2 = 7.9 \times 
10^{-8}$ eV${}^2 $ (LMA-V). The large angle vacuum solution. 
\end{enumerate}

While choosing only these values may appear restrictive, it will
be seen that these values cover the typical ranges within which all the
allowed changes take place. The numerical calculations are done by
following the evolution of the mass eigenstates through all the resonances
including the appropriate jump probabilities.

We will now discuss the results for all these choices. We choose
$\epsilon = 0.087$ for the adiabatic case and $\epsilon \sim 0$ for the
nonadiabatic one. Whether or not the Landau Zener jumps play a role is
determined by the value of $\omega$ we choose. The interaction at the
detector is mainly of three types:
\begin{enumerate}
\item Isotropic events: These are by far the largest fraction of the
events, and are due to $\overline{\nu}_e \, p \to e^+ \, n$.
\item Forward peaked events: These are due to elastic scatterings of
$\nu_e$, $\overline{\nu}_e$, $\nu_x$, $\overline{\nu}_x$ on electron
targets.
\item Backward peaked events: These arise from CC scattering of $\nu_e$
and $\overline{\nu}_e$ on oxygen nuclei in the target.
\end{enumerate}
Hence, angular information on the final state electron ($e^-$ or $e^+$)
will allow us to separate out the above three types of events. Note
however that the extent of backward peaking is severely dependent on the
spectral temperature \cite{haxton}. For the temperatures corresponding
to the $\nu_x$ spectrum, we can assume the events to be totally in the
backward direction.

\subsection{Total Number of events}

We give the time integrated number of events of the scattered electron,
with energy, $E_e > 8$ MeV (which is a typical threshhold for such
detectors), in Tables 4 to 6, for the three choices of $\omega$. These
are the number of events for a supernova explosion at 10 Kpc for a 1
kTon water Cerenkov detector. 

The first column in all the tables corresponds to the base case of no
mixing. The next two columns correspond to the adiabatic result
($\epsilon = 0.087$ for the three and four flavour case, while the last
two columns correspond to the case where $\epsilon \sim 0$. The
isotropic events (due to $\nu_e$) are insensitive to values of
$\epsilon$ chosen. The first table corresponds to small $\omega$, where
the lower resonance can involve Landau Zener jumps, while the next two
correspond to large $\omega$ and are adiabatic at the lower resonance.

It is expected that the integrated events are less sensitive to the
details of models of stellar collapse, especially the energy
distribution of the fluxes. Hence the predictions are more stable.

\subsubsection{$\epsilon = 0.087$ (Adiabatic case)}
The large backward excess due to mixing of a hot spectrum into the
$\overline{\nu}_e$ spectrum is a clear signal of mixing, irrespective of
the number of flavours involved, as seen in Tables 4--6. Variations in the
forward events may not be visible. However, the large depletion in the
isotropic events with increasing $\omega$ (in the 4 flavour case) can
distinguish between the 3 and 4 flavour mixing, especially when $\omega$
is large (see Tables 5 and 6). 

\subsubsection{$\epsilon \sim 0$} Here the lower resonance may be
adiabatic or nonadiabatic, depending on the value of $\omega$. Again
here a large excess of backward events clearly indicates the presence
of mixing, but this is seen only for the 3 flavour case (see Tables 4--6).
The isotropic background from the $\overline{\nu}_e \, p$ scattering is
again sensitive to mixing and the number of mixing flavours, for large
$\omega$ (tables 5 and 6). Finally, unlike the previous case the
depletion (enhancement) of forward events, independent of $\omega$, for
4(3) flavours may be observable, especially in the differential event
rates, $\d N/\d E_e$.  It is therefore interesting to study the
differential spectrum, which obviously has more information than the
integrated events, to see whether this can be quantified further. We do
this below.

\subsection{Event rates}

The event rates as a function of the scattered electron energy, $E_e$,
are displayed in Figs. 3, 4, and 5. For detailed results in the
3-flavour case, see I. The solid lines in all the figures refer to the
case when there is no mixing and serve as a reference. Both the
three- and four-flavour results are displayed in each of these figures.

In Fig. 3 we show the time integrated event rate per unit electron
energy bin (of 1 MeV) as a function of the detected electron energy for
the adiabatic case when $\epsilon = 0.087$. The dashed curves
correspond to the case with mixing. Interactions with both electrons as
well as oxygen nuclei occur.  The event rates in both cases are shown.
In both cases, mixing progressively enhances the high $E_e$ event
rates. This shows up as an increase in both the forward and abckward
peaked events. There is no perceptible difference between the 3- and
4-flavour scenarios, independent of constraints on the angle $\omega$.

In Fig. 4 we show the electron neutrino spectrum for the case when
$\epsilon $ is small, in fact near zero. Hence the propagation near the
upper resonances is nonadiabatic. Here, 3 and 4 flavour mixing give
drastically different results. For all $\omega$, the forward and
backward rates are depleted for 4 flavour mixing. However, these rates
are small. The 3 flavour nonadiabatic mixing is indistinguishable from
the adiabatic case (compare the left hand sides of Figs. 3 and 4). 

Thus, an enhancement in the backward peaked events clearly indicates the
presence of mixing. But there is no unambigious indicator of the number of
flavours unless the parameters get restricted better through solar
neutrino solutions. For this, we need to examine the electron
anti-neutrino spectrum.

In Fig. 5 we show the $\overline\nu_e$ spectrum as a function of energy
for three different choices of $\omega$. (Here $\epsilon = 0.087$ but
the results are essentially the same even if $\epsilon$ is nearly
zero.) The propagation is always adiabatic for the antineutrinos since
there are no resonances to account for nonadiabatic jumps. The
$\overline{\nu}_e$ can interact with electrons, oxygen and protons in
water and therefore contribute to forward, backward and isotropic
events, the most dominant one being the approximately isotropic events
due to interaction on the proton target (this cross-section is about
two orders of magnitude larger than the others, for typical energies
involved).  For the SMA solution, there is no perceptible difference
between the three and four flavour scenarios and the no mixing case, as
already discussed. A perceptible difference occurs for LMA and LMA-V
solutions. The intermediate energy ($E_e \sim 20$ MeV) spectrum is not
depleted for 3 flavour mixing, while it is depleted for 4 flavour
mixing (for the LMA solutions). Again, the 3 flavour isotropic spectrum
is enhanced at larger energies (See Fig.~5); no such enhancement occurs
in 4 flavours. Such an energy dependent pattern of event depletion and
enhancement may help distinguish not only the 3 from the 4 flavour
case, but also the mixing from the no-mixing solution. This pattern
will survive, independent of uncertainties in the overall flux
normalisations; however, it must be remembered that the event rate is
very small at large energies ($E_e > 40$ MeV).

It is therefore seen that a combination of detection of both isotropic
and forward events at low and intermediate energies (involving both
$\overline{\nu}_e\, p$ and $\nu_e\, e$ scattering in water) can help
distinguish the 3 and 4 flavour solutions for the case when $\epsilon$
is close to zero.

\section{Conclusions}
To summarise, neutrino mixing in the 3- and 4- flavour cases gives rise
to very different event rates due to neutrinos (and antineutrinos) from
stellar collapse interacting with a water Cerenkov detector. The
neutrino sector is sensitive to the presence of mixing.  This shows up
typically as an enhancement of the higher energy events in the backward
direction. This signal is not sensitive to the number of flavours. For
a fairly large region of parameter space, the antineutrino events are
sensitive to the number of flavours involved. Since this is the main
signal of stellar collapse, the statistics should be good enough to
make rather strong predictions from this channel.

\bigskip \bigskip
We would like to thank the organisers of the 6th Workshop on High
Energy Physics Phenomenology (WHEPP-6), Chennai, January 2000, where a
part of this work was done.  We thank Sandhya Choubey, Srubabati
Goswami, Kamales Kar, and Debashis Majumdar for discussions during this
meeting. We would also like to thank Hari Dass for discussions.

\section*{Appendix A}
We clarify here our choice of mixing matrix that we have used in the
text. This arises from the requirement of consistency with various
experiments. We discuss this, for the case of vacuum mixing, for both
three and four flavour solutions. We shall set all CP violating phases
to zero in what follows.

For the sake of completeness, we briefly quote the results for the
three flavour case, where all flavours are active neutrinos. We choose
the mixing angles to be $\omega$, $\phi$ and $\psi$ for the (12), (13)
and (23) mixings.  The electron neutrino survival probability is given
by,
$$
\begin{array}{rcl}
P_{ee} & = &  (1-2 c_\phi^2 s_\phi^2) - 2 c_\phi^4 c_\omega^2 s_\omega^2
+ 2 c_\phi^4 c_\omega^2 s_\omega^2 \cos(2.54 \delta_{12}L/E) \\ 
 & & + 
2 c_\phi^4 s_\phi^2 \left[ c_\omega^2 \cos(2.54 \delta_{13}L/E) + 
s_\omega^2 \cos(2.54 \delta_{23}L/E) \right]~,
\end{array}
\eqno(A.1)
$$
where $L$ is measured in meters, $\delta$ in eV${}^2$ and $E$ in MeV.

We first assume a set of weak constraints, viz., that there exist two
different scales for $\delta$: $\delta_{12} \le 10^{-5}$ eV${}^2$ from
solar neutrino data, and $\delta_{13} \sim 10^{-3}$ eV${}^2$ from
atmospheric neutrino data. Hence $\delta_{13} \simeq \delta_{23} =
\delta$.

For the case of {\sc chooz}, we have $L/E \sim 300$ \cite{chooz}.
Then the electron neutrino survival probability that can be applied in
the {\sc chooz} experiment reduces to
$$
P_{ee} = 1 - 4\epsilon^2 \sin^2 \left( 1.27 \frac{\delta L}{E} \right)~,
\eqno(A.2)
$$
independent of $\omega$, where we have set $\sin\phi = \epsilon$. Hence
the {\sc chooz} result constrains $\epsilon$ to be small; in
particular, $\epsilon < 0.16$, which justifies the approximation used
in our parametrisation of the three-flavour mixing matrix.  As a result
one cannot accomodate the {\sc lsnd} results \cite{lsnd} in a
three-flavour frame-work.

Hence it is necessary, for the sake of consistency of atmospheric
neutrinos, {\sc chooz} and {\sc lsnd}, to have at least one extra
(sterile) flavour.

\begin{figure}[htp]
\vskip 8truecm
{\includegraphics{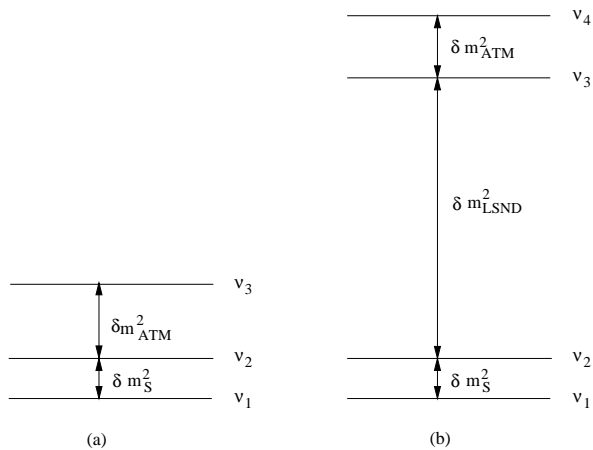}}
\caption{The vacuum mass square differences in the 4 flavour schemes.
Here $\nu_e$ and $\nu_s$ are predominantly mixed states of $\nu_1$ and
$\nu_2$ while $\nu_\mu$ and $\nu_\tau$ are that of $\nu_3$ and $\nu_4$.
The mixing between the lower and upper doublets has been chosen to
be very small. Here S, ATM and LSND stand for the solar, atmospheric and
LSND mass squared differences respectively. We have also shown the
3 flavour scheme for comparison.}
\label{fig:level}
\end{figure}

We now discuss the four-flavour scenario within the mass hierarchy shown in
Figure \ref{fig:level}. The same weak constraints as in the
three-flavour case are assumed here. This results in a two-doublet
structure for the four levels where the doublets are separated by the
mass scale imposed by {\sc lsnd}, namely, $\delta_{13} \simeq \delta_{14}
\simeq \delta_{23} \simeq \delta_{24} \sim 0.1$--1 eV${}^2$. In addition
to these mass squared differences, there are now six mixing angles,
$\theta_{12} = \omega$,
$\theta_{34} = \psi$,
$\theta_{13} = \theta_1$, 
$\theta_{14} = \theta_2$, 
$\theta_{23} = \theta_3$, 
$\theta_{24} = \theta_4$, where $\psi$ is constrained to be maximal,
$\psi = \pi/4$, from atmospheric neutrino data. Hence the mixing matrix is
$$
U = U_\psi U_4 U_3 U_2 U_1 U_\omega~.
\eqno(A.3)
$$
We first discuss the constraint on $P_{ee}$ from {\sc chooz} in the
4-flavour case. Because $L/E$ is large, of the order of 300,
the oscillatory term involving $\delta_{13}$ averages out to zero:
$$
\left\langle \cos \left( \frac{\delta_{13} L}{2E} \right) \right\rangle = 0.
\eqno(A.4)
$$
The same holds for other mass squared differences of the same order of
magnitude as $\delta_{13}$.  Furthermore, since $\delta_{12}$ is small,
we we can set $\cos (\delta_{12} L/(2E)) \sim 1$. Therefore the
survival probability of the electron neutrino is,
$$
P_{ee} = 1 - \frac12 \sin^2 2\theta_2 - \frac12 \cos^4 \theta_2 \sin^2 2
\theta_1 + \frac12 \sin^2 2\theta_2 \sin^2 2 \theta_1 \cos \left(
\frac{\delta_{34}L}{2E} \right)~,
\eqno(A.5)
$$
where the scale $\delta_{34}$ is set by the atmospheric neutrino data so
that the oscillatory term is of order one. Note that the survival
probability does not involve $\theta_3$ and $\theta_4$, as remarked in
the text.

Then the {\sc chooz} result, $P_{ee} = 1 \pm 0.04$ \cite{chooz} indicates
that both $\theta_1$ and $\theta_2$ should be small. We have therefore chosen
these to be of the same order: $\sin\theta_1 \sim \sin\theta_2 \sim
\epsilon$. Then
$$
\begin{array}{rcl}
P_{ee} & = &  1 - 4 \epsilon^2 + 2 \epsilon^4 \cos\left(
			\frac{\delta_{34}L}{2E} \right)~, \\
       & \simeq & 1 - 4\epsilon^2~,
\end{array}
\eqno(A.6)
$$
independent of $\delta_{34}$, $L$, and $E$. The {\sc chooz} null
result therefore provides an upper bound on these mixing angles.

Using all these constraints, the relevant
transition probability for {\sc lsnd} is given by
$$
P_{\mu e} = 2 \epsilon^2 \left[ \cos\theta_3 - \sin\theta_4 \sin\theta_3
		       + \cos\theta_4 \right]^2 
		       \sin^2 \left( \frac{\delta L}{4E} \right)~.
\eqno(A.7)
$$
Here $\delta$ is the separation between the two doublets.

The {\sc lsnd} result gives a non-zero value for $P_{\mu e}$, viz.,
$P_{\mu e} = (2.6 \pm 1.1) \times 10^{-3}$ \cite{lsnd}. This
unambiguously fixes $\epsilon$ to be different from zero even though
the {\sc chooz} result allows this value. Combining the limits from the
two experiments, we have the constraint,
$$
\epsilon_{\rm{LSND}} \le \epsilon \le \epsilon_{\rm{CHOOZ}}~.
\eqno(A.8)
$$
The values we get in the four-flavour case are a completely consistent
set. We have used allowed values of the various parameters from the above
in the analysis of supernova neutrinos in the text. 

Note that there are still no constraints on $\theta_{3}$ and
$\theta_{4}$ which are the (23) and (24) mixing angles. Although we
have set them to $\epsilon$ in the text, the electron (and
anti-electron) survival probabilities that we define are independent of
these mixing angles (since they do not involve electron-type
transitions). Hence this choice does not affect the results that we
have obtained for these probabilities. Furthermore, we remark that the
present generation of experiments is unlikely to constrain these angles
significantly. A more detailed analysis of the allowed parameter space
in the case of four-flavour mixing will be published elsewhere.

\newpage

\begin{table}[htb]
\centering
\begin{tabular}{|c|l|} \hline
No. of flavours & Neutrino flux at detector, $F_f$ \\ \hline
3 & $F_{\nu_e} =\epsilon^2 F_{\nu_e}^0 + (1 - \epsilon^2) F_{x}^0$ \\
4 & $F_{\nu_e} =\epsilon^2 F_{\nu_e}^0 + (1 - 2\epsilon^2) F_{x}^0$ \\
\hline
3 & $2 F_{x} = (1+\epsilon^2) F_{x}^0 + (1-\epsilon^2) F_{\nu_e}^0$ \\
4 & $2 F_{x} = (4\epsilon^2)F_{x}^0 + (1-2\epsilon^2) F_{\nu_e}^0$ \\ \hline
4 & $F_s =  (1-2\epsilon^2) F_{x}^0 +
               (\epsilon^2) F_{\nu_e}^0$ \\ \hline
\end{tabular}
\caption{Neutrino fluxes observed at the detector in
the extreme adiabatic limit.}
\label{tab:adnu}
\end{table}

\begin{table}[htb]
\centering
\begin{tabular}{|c|l|} \hline
No. of flavours & {Antineutrino flux at detector, $F_f$} \\ \hline
3 & $F_{\bar{e}} = (1-\epsilon^2)c^2_\omega F_{\bar{e}}^0
+ (s^2_\omega + \epsilon^2 c^2_\omega) F_{x}^0$ \\
4 & $F_{\bar{e}} = (1-2\epsilon^2)c^2_\omega F_{\bar{e}}^0
+ (2\epsilon^2) F_{x}^0$ \\ \hline
3 & 2$F_{\bar{x}} = (1+c^2_\omega -\epsilon^2c^2_\omega)
F_{x}^0 +(s^2_\omega +\epsilon^2c^2_\omega)
F_{\bar{e}}$ \\
4 & 2$F_{\bar{x}} = (2 -4\epsilon^2) F_{x}^0 +
+2\epsilon^2(1-s_{2\omega}) F_{\bar{e}}$ \\ \hline
4 & $F_{\bar{s}} = 2\epsilon^2 F_{x}^0 + 
(s^2_\omega -2\epsilon^2(s^2_\omega-s_{2\omega}))
F_{\bar{e}}$ \\ \hline
\end{tabular}
\caption{Antineutrino fluxes at the detector. $c_\omega = \cos\omega$,
$s_\omega = \sin \omega$.} 
\label{tab:adnubar}
\end{table}

\vspace{1cm}

\begin{table}[ht]
\centering
\begin{tabular}{|c|l|} \hline
No. of flavours & Neutrino flux at detector, $F_f$ \\ \hline
3 & $F_{e} =(1-\epsilon^2)[ (1-P_L)s_{\omega}^2 + P_L c_\omega^2] 
F_{e}^0$ \\
 & $+
[1- (1-\epsilon^2)((1-P_L)s^2_{\omega} + P_Lc_{\omega}^2)]F_{x}^0$ \\
4 & $F_{e} =(1-2\epsilon^2)[ (1-P_L)s_{\omega}^2 + P_L c_\omega^2] 
F_{e}^0 +
2\epsilon^2 F_{x}^0$ \\ \hline

3 & $2F_{x} =[1+(1-\epsilon^2)( (1-P_L)s_{\omega}^2 + P_L c_\omega^2)] 
F_{x}^0$ \\
  & $ +
[1- (1-\epsilon^2)((1-P_L)s^2_{\omega} + P_Lc_{\omega}^2)]F_{e}^0$ \\
4 & $2F_{x} =2(1-2\epsilon^2)F_{x}^0 +
2\epsilon^2 \left[1+(1-2P_L)s_(2\omega)\right] F_{e}^0$ \\ \hline

4 & $F_s =  2\epsilon^2F_{x}^0 +
\left\{ [(1-P_L)c^2_{\omega} + P_L s_{\omega}^2] -
2\epsilon^2 \left[c^2_{\omega} - s_{\omega}^2 - P_L [
c_{2\omega} + 2 s_{2\omega}] \right] \right\} F_e^0$ \\
\hline
\end{tabular}
\caption{Neutrino fluxes at the detector when non-adiabatic 
effects are introduced. Note that while the transition is assumed to be 
fully non-adiabatic at the upper resonances, it is controlled by the 
probability $P_L$ at the lower resonance.} \label{tab:nonad} 
\end{table}

\newpage

\begin{table}[htp]
\begin{tabular}{|r|r|r|r||r|r|}
\hline
Events & No Mixing & \multicolumn{2}{c||}{Mixing$(\epsilon=0.087)$} 
       & \multicolumn{2}{c|}{Mixing$(\epsilon \sim 0)$}\\ 
   & & 3 flavour & 4 flavour & 3 flavour & 4 flavour \\ \hline
Forward & 5.9 & 7.6 & 6.9 & 7.1 & 4.4\\ \hline
Backward & 4.8 & 27.3 & 27.1 & 21.5 & 4.2\\ \hline
Isotropic & 281.1 & 282.0 & 282.2 & 281.3 & 280.7 \\ \hline
\end{tabular}
\caption {Total number of events (for two different choices of $\epsilon$)
in the forward and backward direction,
and isotropic events, respectively, for a scattered electron energy
above 8 MeV. Here $\sin^2(2\omega)=6\times 10^{-3}, 
         \delta^2 m_{12}=5.4 \times 10^-6$ eV$^2$.}   \label{sma}
\end{table}

\begin{table}[htp]
\begin{tabular}{|r|r|r|r||r|r|}
\hline
Events & No Mixing & \multicolumn{2}{c||}{Mixing$(\epsilon=0.087)$} 
       & \multicolumn{2}{c|}{Mixing$(\epsilon \sim 0)$}\\ 
   & & 3 flavour & 4 flavour & 3 flavour & 4 flavour \\ \hline
Forward & 5.9 & 7.6 & 6.6 & 7.2 & 3.8 \\ \hline
Backward & 4.8 & 29.5 & 26.2 & 23.8 & 3.1 \\ \hline
Isotropic & 281.1 & 306.7 & 212.0 & 306.1 & 209.4\\ \hline
\end{tabular}
\caption {Same as above for  
   $\sin^2(2\omega)=0.76, \delta^2 m_{12}=1.8 \times 10^-5$ eV$^2$.}
						\label{lma}
\end{table}

\begin{table}[htp]
\begin{tabular}{|r|r|r|r||r|r|}
\hline
Events & No Mixing & \multicolumn{2}{c||}{Mixing$(\epsilon=0.087)$} 
       & \multicolumn{2}{c|}{Mixing$(\epsilon \sim 0)$}\\ 
   & & 3 flavour & 4 flavour & 3 flavour & 4 flavour \\ \hline
Forward & 5.9 & 7.7 & 6.5 & 7.0 & 4.0 \\ \hline
Backward & 4.8 & 30.7 & 25.6 & 21.8 & 2.7 \\ \hline
Isotropic & 281.1 & 320.8 & 171.9 & 320.3 & 168.7\\ \hline
\end{tabular}
\caption { Same as above for $\sin^2(2\omega)=0.96, 
         \delta^2 m_{12}=7.9 \times 10^-8$ eV$^2$.}      \label{lmav}
\end{table}

\begin{figure}[htp]
\vskip 8truecm
{\includegraphics{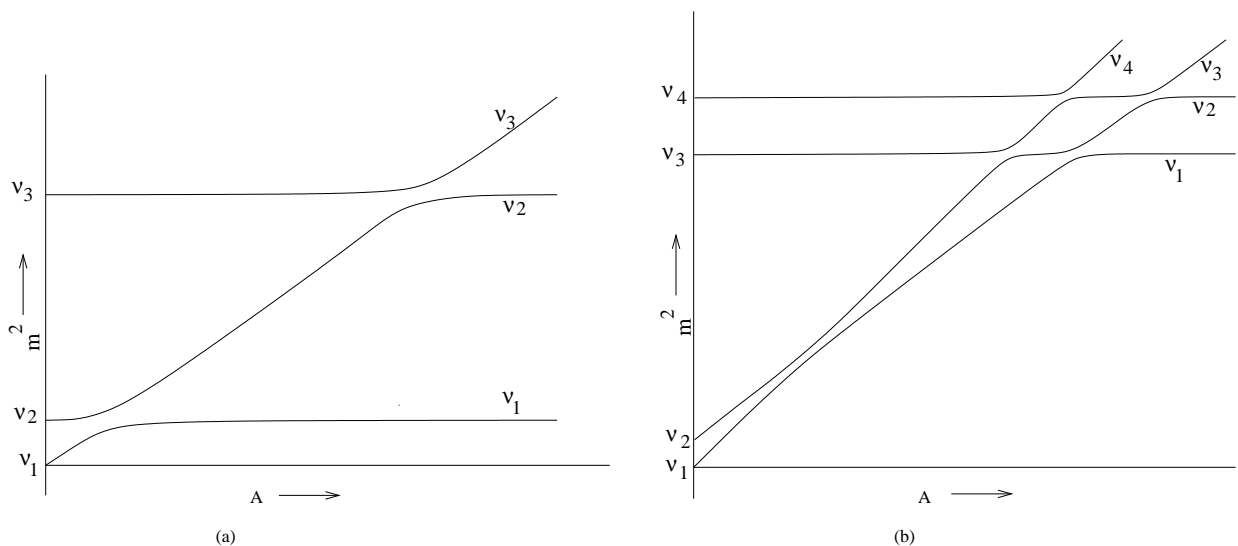}}
\caption{Schematic drawing showing mass squares as functions of matter
density in the 4 flavour scheme.  Resonances occur at two different
regions of matter density, the lower one at $A\approx \delta m^2_S$ and
the upper one at $A\approx \delta m^2_{\rm LSND}$. The upper one
consists of 4 close resonances. The three flavour scheme is also shown
for comparison.}
\end{figure}

\newpage
~ 
\begin{figure}[htp]
\vskip 9truecm
{\includegraphics{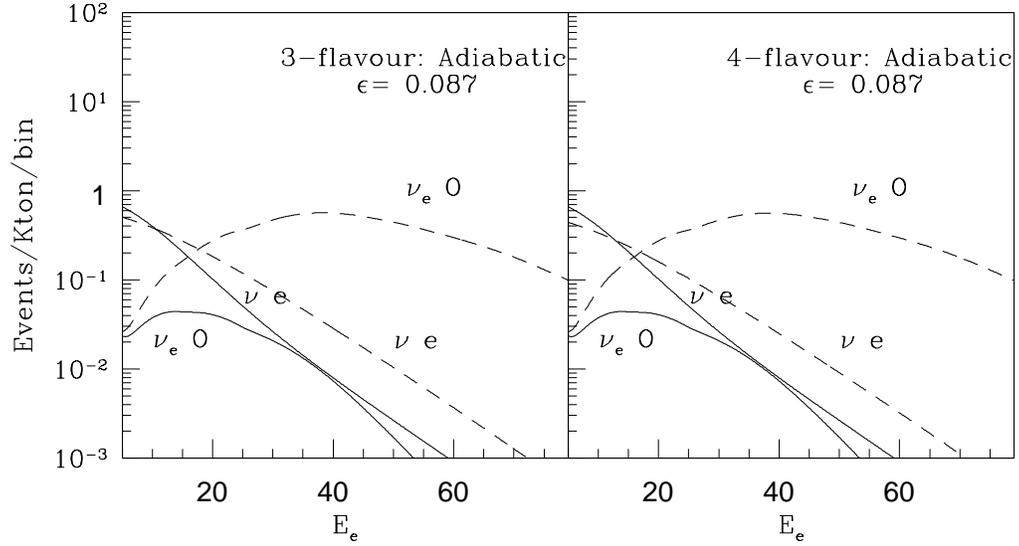}}
\caption{$\nu_e\, O$ and $\nu \, e$ (for all flavours of $\nu$) event
rates when the upper resonance is completely adiabatic. The solid lines
represent the no mixing case and is plotted in all the graphs for
comparison. The dashed lines are due to the effects of mixing. The
oxygen events show dramatic increase due to mixing.  Note that 3 and 4
flavour cases cannot be distinguished.}
\end{figure}

\newpage
~
\begin{figure}[htp]
\vskip 18truecm
{\includegraphics{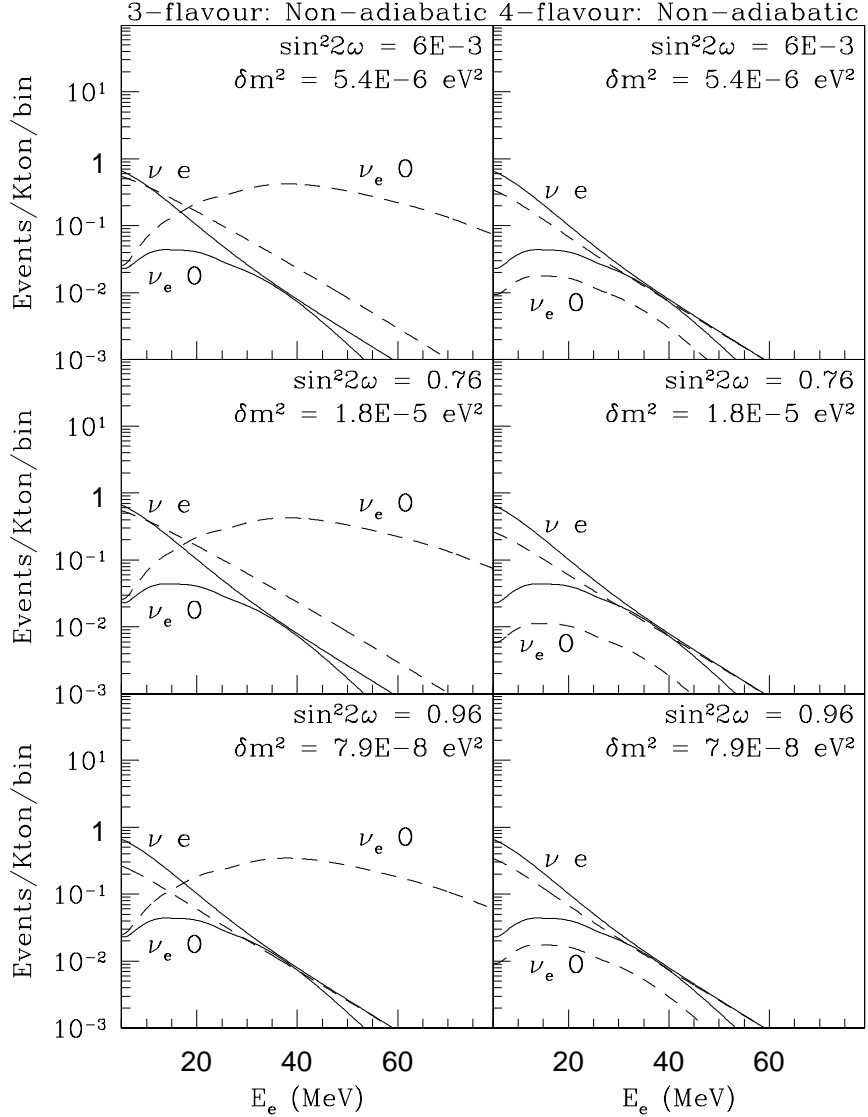}}
\caption{$\nu_e\, O$ and $\nu \, e$ (for all flavours of $\nu$) event
rates when the upper resonance is completely non-adiabatic. The results
depend upon the three possible solutions to the solar neutrino puzzle
and are shown in the three panels, top, middle and bottom. The three
flavour results are similar to the adiabatic case shown in Fig.\ 3 but
the 4 flavour case shows suppression of the event rates in all cases.
Here the different cases are distinguished by the extent of
suppression.}
\end{figure}

\newpage
~
\begin{figure}[htp]
\vskip 18truecm
{\includegraphics{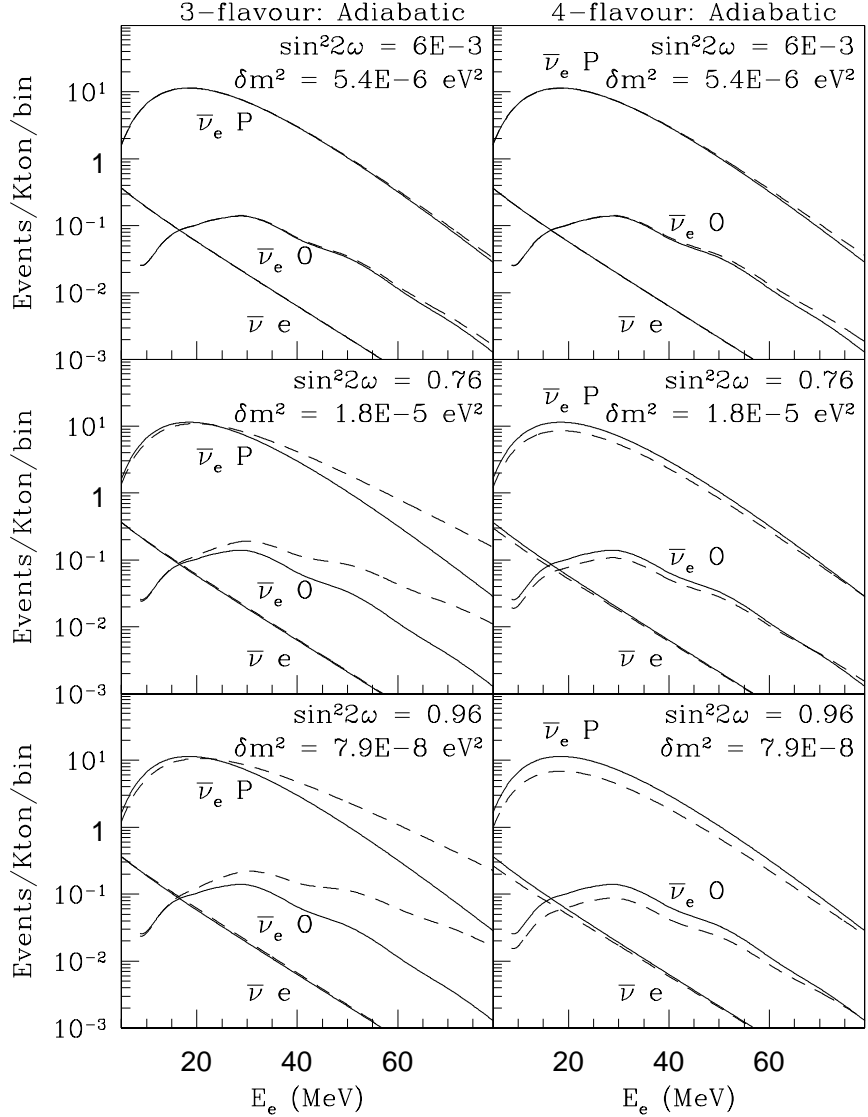}}
\caption{$\bar{\nu_e}$ event rates for $\epsilon = 0.087$. The solid
(dashed) lines are due to (no) mixing. While the 3 flavour scheme shows
enhancement of event rates at high energies due to mixing, the 4
flavour scheme shows suppression at lower energies.}
\end{figure}

\end{document}